\documentclass[journal=mamobx,manuscript=article]{achemso}

\usepackage{graphicx}  
\usepackage{bm} 
\usepackage{amssymb} 
\usepackage{color}

\newcommand{\eps}[0]{\varepsilon}
\newcommand{\deps}[0]{\dot{\varepsilon}}
\newcommand{\degC}[0]{$^{\circ}$C}
\newcommand{\w}[0]{\omega}

\title{Nonlinear Viscoelastic Modeling of Adhesive Failure for Polyacrylate Pressure-Sensitive Adhesives}

\author{Julien Chopin}
\affiliation[ESPCI]
{Laboratoire Sciences et Ing\'enierie de la Mati\`ere Molle, PSL Research University, UPMC Univ Paris 06, ESPCI Paris, CNRS, 10 rue Vauquelin, 75231 Paris cedex 05, France}
\alsoaffiliation[Salvador]
{Instituto de F\'isica,  Universidade Federal da Bahia, Salvador-BA 40170-115, Brazil}
\email{julien.chopin@ufba.br}
\author{Richard Villey}
\affiliation[Saint Gobain]
{Saint-Gobain Glass France, Chantereine R\&D Center, 1 rue de Montlu{\c c}on - BP 40103, 60777 Thourotte Cedex, France,}
\author{David Yarusso}
\affiliation[MMM]
{3M Center, 3M Company, 230-1D-15, St. Paul, MN, 55144-1000, USA}
\author{Etienne Barthel}
\alsoaffiliation[ESPCI]
{Laboratoire Sciences et Ing\'enierie de la Mati\`ere Molle, PSL Research University, UPMC Univ Paris 06, ESPCI Paris, CNRS, 10 rue Vauquelin, 75231 Paris cedex 05, France}
\author{Costantino Creton}
\alsoaffiliation[ESPCI]
{Laboratoire Sciences et Ing\'enierie de la Mati\`ere Molle, PSL Research University, UPMC Univ Paris 06, ESPCI Paris, CNRS, 10 rue Vauquelin, 75231 Paris cedex 05, France}
\author{Matteo Ciccotti}
\alsoaffiliation[ESPCI]
{Laboratoire Sciences et Ing\'enierie de la Mati\`ere Molle, PSL Research University, UPMC Univ Paris 06, ESPCI Paris, CNRS, 10 rue Vauquelin, 75231 Paris cedex 05, France}
\email{matteo.ciccotti@espci.fr}
\keywords{Adhesion, Cross-linked polymers, peeling, Viscoelasticity, Finite deformations}
\begin{document}

\begin{abstract}
We investigate experimentally the adherence energy $\Gamma$ of model polyacrylate Pressure Sensitive Adhesives (PSAs) with combined large strain rheological measurements in uniaxial extension and an instrumented peel test. We develop a nonlinear model for such peel test which captures the dependence of $\Gamma(V)$ with peeling rate $V$ revealing the key role played by the extensional rheology. Our model explains in particular why traditional linear viscoelastic approaches correctly predict the slope of $\Gamma(V)$ curves for sufficiently elastic PSAs characterized by a simple rate-independent debonding criterion. However, for more viscoelastic adhesives, we identified a more complex rate-dependent debonding criterion yielding a significant modification of the $\Gamma(V)$ curves, an effect that has been largely overlooked so far. This investigation opens the way towards the understanding of fibrils debonding, which is the main missing block to predict the adherence of PSAs.
\end{abstract}

\section{Introduction}
It is a very well known result that during the peeling of a typical viscoelastic PSA from a solid substrate, the measured adherence energy $\Gamma$ (the energy required to peel a unit tape area) is several orders of magnitude above the value given by the thermodynamic work of adhesion between the adhesive and the substrate. Moreover, in the slow steady state regime, $\Gamma$ is very dependent on the temperature $T$ and peeling rate $V$. It has long been observed that for viscoelastic adhesives the adherence curves $\Gamma(V,T)$ can be collapsed to a single master curve $\Gamma(a_T V,T^{ref})$ at a reference temperature $T^{ref}$ by renormalizing the velocity axis by the same shift factor $a_T$ that is used for the Time-Temperature Superposition (TTS) of the linear rheological measurements~\cite{Ferry1970,Gent1969,Kaelble1964,andrews1974mechanics,Maugis1985}. This has suggested that linear viscoelasticity should be used to quantitatively model the adherence energy $\Gamma(V,T)$.

Various linear viscoelastic models have since been developed based on a perturbation of the linear elastic fracture mechanics (LEFM) crack tip singular fields \cite{mueller1971crack,Schapery1975,deGennes1988,Saulnier2004,Persson2005,Barthel2009}. These models predict a fracture energy in the separable form $\Gamma(V,T) = \Gamma_0 \left( 1 + \Phi (a_T V) \right)$ \cite{Gent1972,Maugis1988}, where $\Gamma_0$ is an intrinsic adhesion energy and $\Phi (a_T V)$ is a factor accounting for the linear viscoelastic losses. However, most attempts to quantitatively check these predictions experimentally for soft adhesives and rubbers have up to now failed \cite{Gent1996,Barthel2009,Cristiano2011}. Even an important protagonist of this domain such as Gent~\cite{Gent1996}, has pointed out the intrinsic limitation of these theories to describe experimental data since they would predict a process-zone of unphysical subatomic size.

When considering soft PSAs the elastoadhesive lengthscale $\Gamma/E$, where large strains are present around the crack tip \cite{Hui2003,Creton2016}, is comfortably larger than the typical 20 $\mu$m thickness of the adhesive layer used for tapes \cite{Villey2015}. Since the LEFM approach relies on a clear separability of scales between the crack tip singular field and the bulk sample deformation, it can not be applied to PSAs in most practical situations. Alternative modeling approaches have been developed to account for the strong confinement of the material based on a linear viscoelastic foundation following the pioneering work of Kaelble~\cite{Kaelble1959,Kaelble1960,Kaelble1964}. But the experimentally observed development of a dense array of highly stretched fibrils in the debonding region is at odds with this linear approach~\cite{urahama1989effect,Chiche2005,benyahia1997mechanisms,Gay1999}. The role played by the fibrillated region in the overall dissipation was pointed out by several polymer rheologists~\cite{Gent1969,Derail1997,Yarusso1999}, but most experimental investigations were made on non-crosslinked fluid-like adhesives, where failure occurs either cohesively in the fibrils or without fibrillation at all. Such viscoelastic fluids even if able to strain harden in extension are not representative of the lightly crosslinked polymers used for PSA that form an extended fibrillar zone that cleanly debonds from the surface of the substrate. Thus, the respective roles played in the peeling process by the linear, small strain dissipation controlled by the monomer friction coefficient and by the non-linear extensional rheology, where strain hardening due to chain architecture is important, remains elusive. The same conclusion has been reached with other classically used adhesive tests such as the probe-tack test~\cite{Deplace2009deformation,creton2009large}.

In a previous investigation~\cite{Villey2015}, we specifically studied a series of model lightly cross-linked polyacrylate adhesives with different behaviors in extensional rheology but nearly identical small  strain dissipation at the strain rates relevant for peeling.  We showed from steady-state peeling experiments that a more pronounced strain hardening caused an overall decrease of the peeling force and a stronger dependence with peeling rate, which can not be explained by any model based on linear rheology. In parallel, a quantitative characterization of the debonding region by optical microscopy revealed that the fibrils always detached cleanly from the substrate after a stretch larger than 500\%, which diminished with peeling velocity, confirming that non-linear material properties matter here. Similar experiments were later performed and reported by Barrios \cite{Barrios2017}.
Moreover, we found that a more pronounced strain stiffening causes the maximum fibril extension before detachment to decrease and to become less sensitive to the peeling rate~\cite{Villey2015}.

In this paper, we provide the first quantitative physically based model connecting non-linear extensional rheological properties of model crosslinked PSAs with the adherence curves obtained in peel tests. We reconcile the TTS principle observed in the adherence curve and nonlinear deformation in the adhesive. Our results reveal complex debonding criteria, which unambiguously demonstrate the non trivial role played by polymer chain architecture and hence non-linear rheology in determining the $\Gamma(V)$ curves.\\

\section{Experimental}
The model PSA were synthesized by free radical polymerization in solution from 85 wt\% of 2-ethylhexyl acrylate (EHA), 10 wt\% of methyl acrylate (MA) and 5 wt\% of acrylic acid (AA).  0.2 wt\% or 0.4 wt\% of aluminium acetyl acetonate relative to monomer was added after synthesis to provide two levels of cross-linking (labelled A, B) to the adhesive layer-which were coated between two silicone release liners.

\begin{figure*}[t]
    \centering
    \includegraphics[width=12cm]{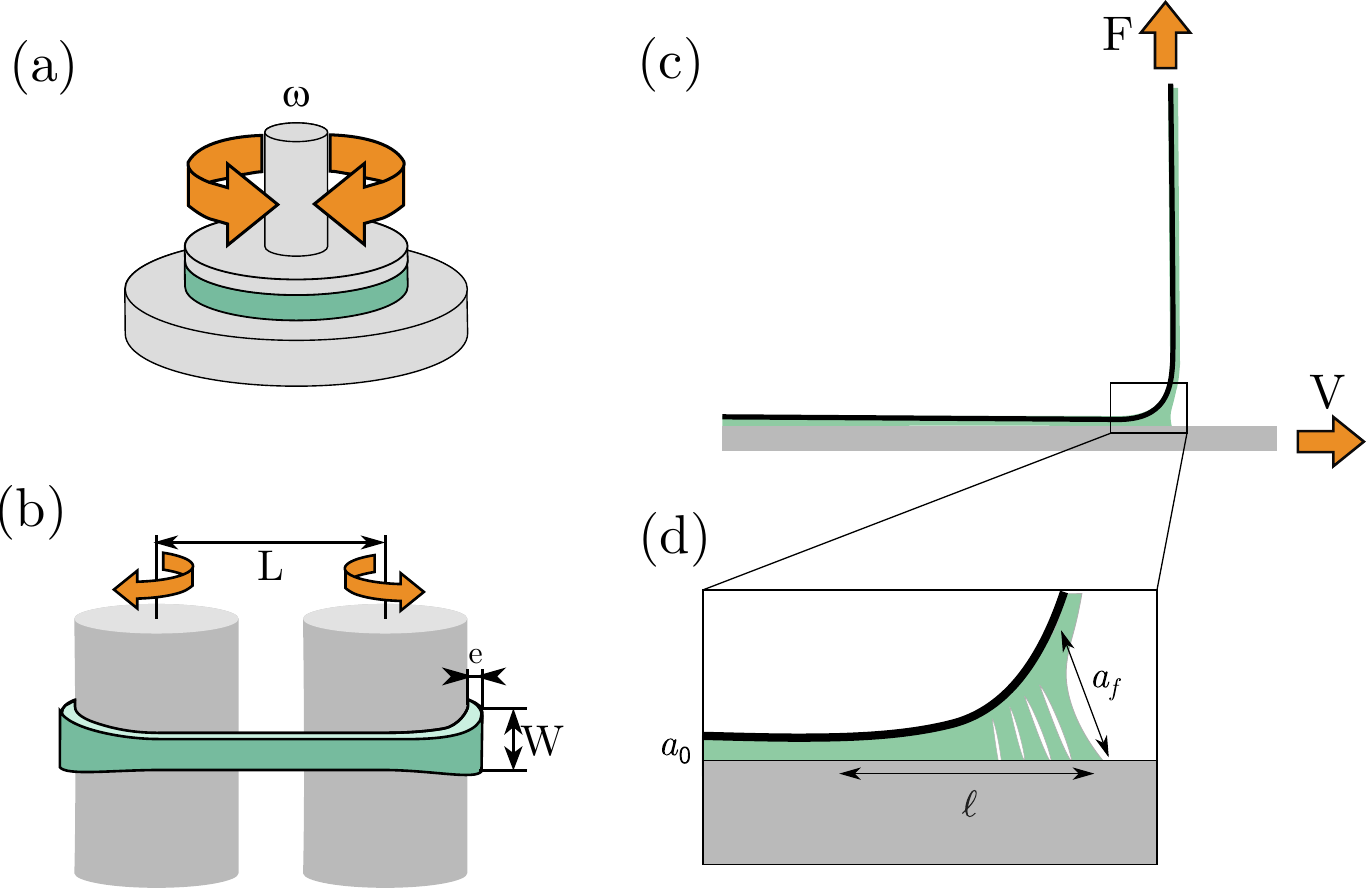}
    \caption{Schematics of experimental setups. (a) The linear shear storage $G'(\w)$ and loss $G''(\w)$ moduli of the adhesives are measured by dynamical mechanical analysis in a parallel plate geometry at imposed pulsation $\w$. (b) Nonlinear rheological properties are measured with the Sentmanat extensional rheometer which measures the applied force on an adhesive slab under uniaxial stretching condition at constant strain rate. (c) Instrumented peeling tests are performed with the same adhesives coated on a PET backing which is peeled by a vertical force $F$ and peeling rate $V$. (d) Sketch of the fibrillar debonding region where $\ell$ is the lateral size and $a_f$ the maximal fibril size before detachment from the substrate.
    }
    \label{Fig-Setups}
\end{figure*}

The linear viscoelastic characterization of the polymers was carried out with an Anton-Paar rheometer (MCR-301) using a standard parallel plate geometry (Fig.~\ref{Fig-Setups}(a)). The storage $G'(\omega,T)$ and loss $G''(\omega,T)$ shear moduli were measured at pulsation $\omega$ in the range $2\cdot 10^{-1}$ to $60$\,rad/s. Pulsation sweeps were performed for temperatures $T$ in the range -40 to 120\degC. The maximum applied shear strain is $\gamma = 2 \cdot 10^{-2}$.

The nonlinear rheological properties are obtained by extensional rheology measurements using the Sentmanat extensional rheometer (SER, see Fig.~\ref{Fig-Setups}(b))~\cite{sentmanat2005measuring}, adapted on the same MCR-301 rheometer. A slab of the polymer of width $W = 5$\,mm and thickness $e = 600\,\mu$m is brought in contact with the cylinders. The stickiness of the materials guarantees a no slip condition at the interface.  The cylinders were rotated at a constant angular velocity, resulting in a constant true strain rate $\dot{\varepsilon}$ in the range $10^{-2}$ to  $4$\,s$^{-1}$. The temperature was varied in the range $0$ to $40$\degC. The cross section is given by $A(t) = e W \cdot \exp (-\dot{\varepsilon} t)$. True strain and stress are given by $\varepsilon(t)=\deps t$ and $\sigma(t) = F(t)/A(t)$, respectively, where $t$ is time and $F$ is the measured force acting on the slab.  Nominal stress is $\sigma_N = \sigma/\lambda$, where $\lambda = \exp(\eps)$ is the stretch.

Finally, instrumented peel tests were performed at a 90$^o$ angle by applying a force $F$ at a peeling rate $V$ (Fig.~\ref{Fig-Setups}(c)). As detailed elsewhere~\cite{Villey2015}, the peeling occurred either at imposed load or imposed velocity allowing to explore a large range of peeling rates from $1\,\mu$m/s to $4\,$m/s. The same model adhesives were coated on a 38$\mu$m thick, PET film. The adhesive layer thickness $a_0 = 19\,\mu$m. All adhesive tapes were carefully bonded to the release side of a commercially available adhesive tape to ensure clean peeling and repeatable results. In this geometry, the energy release rate $\mathcal{G}$ is simply given by :
\begin{equation}
    \mathcal{G} = \frac{F}{b}\,,
\end{equation}
where $b = 20\,$mm is the tape width. In steady state, an energy balance yields $\mathcal{G} = \Gamma(V)$. Further, the experimental setup was equipped with a lateral optical microscope allowing to image the structure of the debonding region with a micrometric resolution (Fig.~\ref{Fig-Setups}(d)). 

Thus, the peeling test was intrumented to characterize the structure of the debonding region while measuring macroscopically the adherence curves $\Gamma(V)$. Insights of viscoelastic dissipative processes occurring in the debonding region have been obtained from linear and nonlinear rheological measurements in controlled and imposed geometries. In the remaining part of the paper, we will present in turn the results obtained from each experimental setup. We will finally present our model, developed to predict the macroscopic adherence curves of lightly crosslinked PSA from bulk rheological properties and detailed knowledge of the structure of the debonding region.

\section{Results}
\begin{figure*}[t]
    \centering
    \includegraphics[width = 16cm]{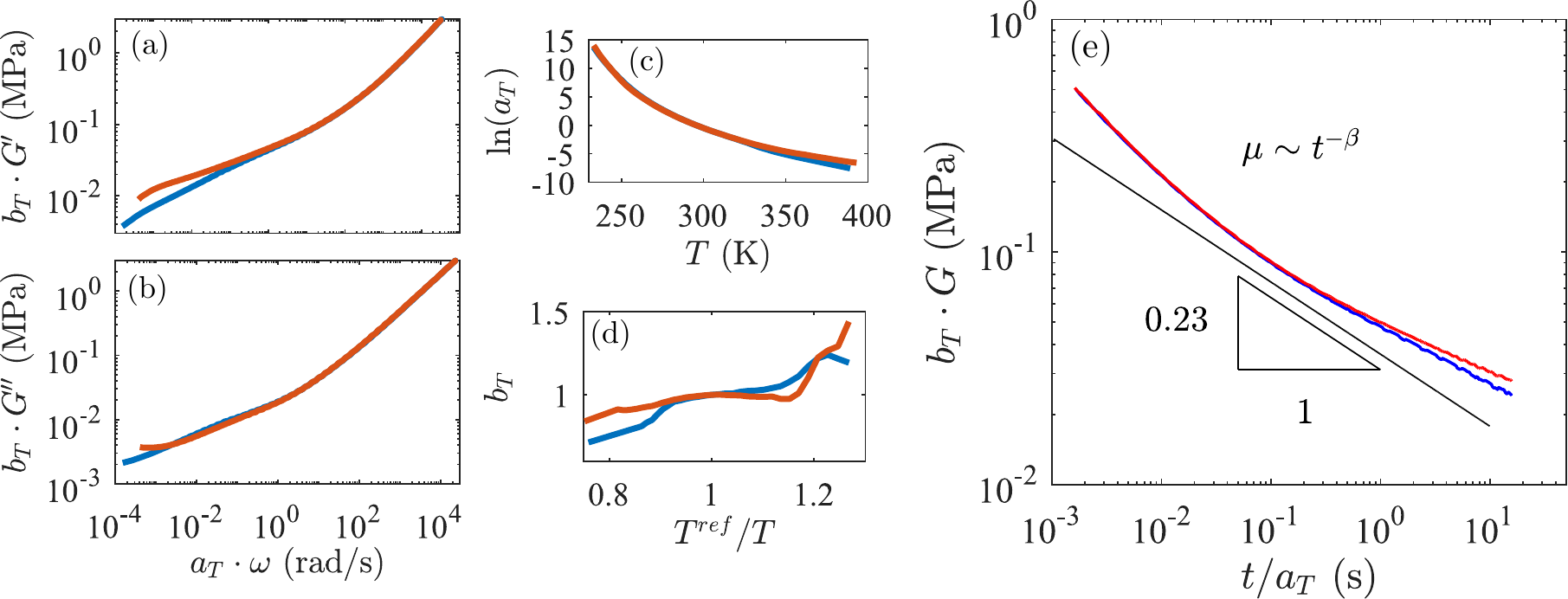}
    \caption{(a) Rescaled storage shear modulus $b_T\cdot G'$ and (b) loss shear modulus $b_T\cdot G''$ for both polymers, which present an identical linear rheological response for $a_T\cdot\omega > 10^{-1}$\, (rad/s). Reference temperature is  $T^{ref} = 296$\,K. (c) Horizontal shift factor $a_T$ and (d) vertical shift factor $b_T$ as a function of temperature $T$ for both polymers. (e) Rescaled relaxation shear modulus $b_T \cdot G(t)$. Its time dependence is roughly approximated by $G \sim t^{-\beta}$ with $\beta = 0.23$.}
    \label{Fig-LinRheol}
\end{figure*}

In Fig.~\ref{Fig-LinRheol}, we present the results obtained from linear rheological measurements.  We plot the rescaled storage shear modulus $b_T\cdot G'$ in Fig.~\ref{Fig-LinRheol}(a) and the rescaled loss modulus $b_T\cdot G''$ in Fig.~\ref{Fig-LinRheol}(b) as a function of the rescaled pulsation $a_T\cdot \omega$ at a reference temperature $T^{ref} = 296$\,K. The horizontal shift factor $a_T$ and vertical shift factor $b_T$ are shown in Fig.~\ref{Fig-LinRheol}(c) and (d), respectively. Over the range of temperatures and rates used in the peel tests and extensional rheology, $b_T \approx 1$ within 10\% error. The dependence of the horizontal shift factor with temperature can be captured using the Williams-Landel-Ferry model :
\begin{equation}
    \ln a_T = -C_1 \frac{T-T^{ref}}{C_2+T-T^{ref}}
\end{equation}
with $T^{ref} = 296$\,K. A fit of the experimental data over the entire temperature range gives for polymer A: $C_1 = 19.5$ and $C_2 = 154.4$\,K, and, for polymer B: $C_1 = 12.6$ and $C_2 = 118.6$\,K. In Fig.~\ref{Fig-LinRheol}(e), we calculate the relaxation shear modulus $G(t)$ at temperature $T^{ref}$ using the Cox-Merz rule, yielding $G\left( t = \omega ^{-1}\right) \simeq \sqrt{G'^2(\omega)+G''^2(\omega)}$. This  can be roughly approximated by a power law $G \sim t^{-\beta}$ with $\beta = 0.23$ in the relevant range for peel. We can readily see that the level of cross-linking does not significantly affect the linear rheological response of the material nor the shift factors.\\

\begin{figure*}[t]
    \centering
    \includegraphics[width = 12cm]{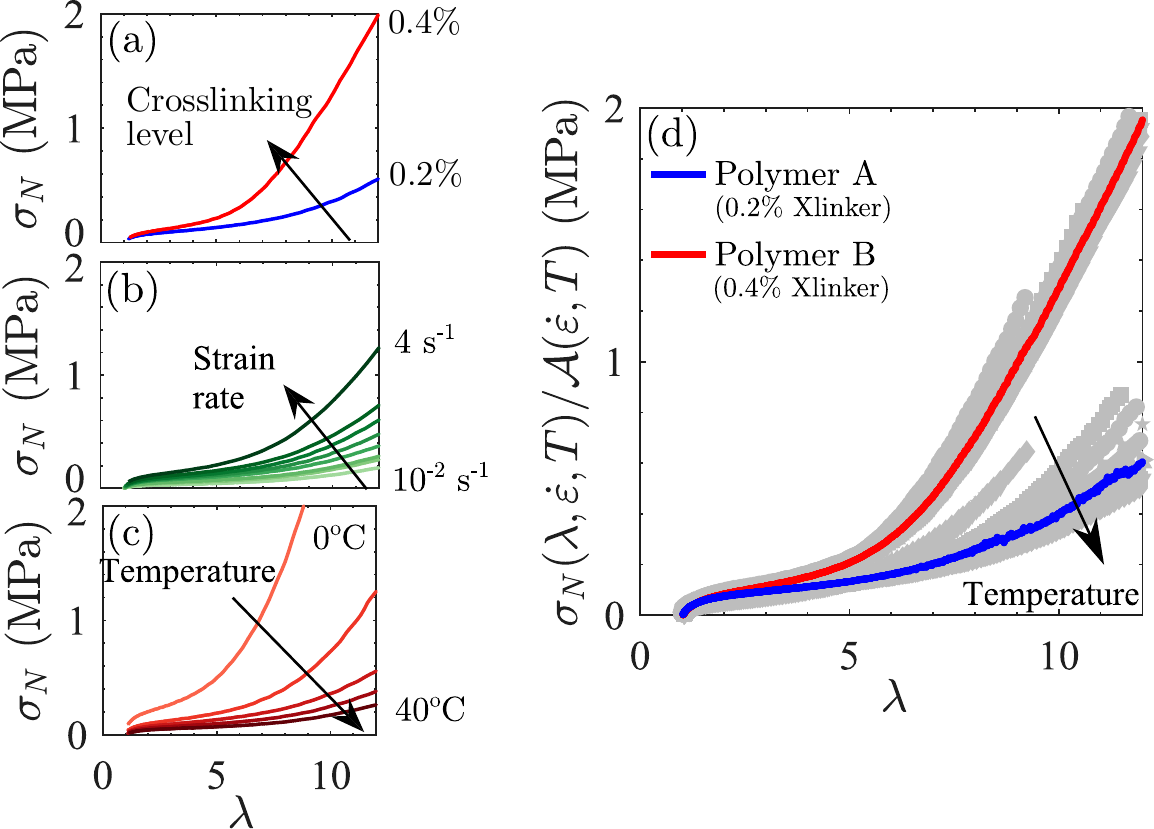}
    \caption{(a-c) Nonlinear extensional curves representing the nominal stress $\sigma_N$ versus stretch $\lambda$, varying (a) concentration of crosslinker level, (b) strain rate $\deps$, and (c) temperature $T$. All the curves show a softening regime followed by a strain hardening. Higher concentrations of crosslinker induce stiffening at large $\lambda$ but makes no observable difference at small $\lambda$.  (d) All the rheological curves collapse on a master curve when rescaled by a single paremeter $\mathcal{A}(\deps,T)$ independent of $\lambda$.
    }
    \label{Fig-NonLinExt}
\end{figure*}

Typical nonlinear  extensional  curves $\sigma_N(\lambda)$ are shown in Fig.~\ref{Fig-NonLinExt}(a-c) for independent changes of  the crosslinking level, strain rate and temperature. In all cases, after a linear regime at $\lambda  \sim 1$, we observe a softening followed by a strain hardening, as expected for PSAs~\cite{deplace2009fine}. 
Fig.~\ref{Fig-NonLinExt}(a) shows that when increasing the level of crosslinking (at $\deps = 1$\,s$^{-1}$ and $T = 20$\degC) the curves are first superimposed in the linear regime as expected and then significantly depart from it at $\lambda \approx 3-4$ while still in the softening regime. The typical stretch $\lambda_c$ for the onset of strain hardening is $\lambda_c = 7$ (resp.\ 5) for polymer A \textcolor{black}{(resp.\ B)}. Fig.~\ref{Fig-NonLinExt}(d) shows that the variation of $\sigma_N$ with $\deps$ and $T$ can be captured by a single prefactor $\mathcal{A}(\deps,T)$ allowing to collapse all the curves on a reference master curve, where $\mathcal{A}(\deps^{ref}=1$\,s$^{-1},T^{ref} = 296\textrm{K}) = 1$. The stress curves can then be written in a separable form $\sigma_N(\lambda,\deps,T) = \mathcal{A}(\deps,T)\cdot \sigma^{ref}_N(\lambda)$, where $\sigma_N^{ref}(\lambda)$ only depends on the stretch $\lambda$ and on the cross-linking level. This separability is less obvious for the polymer A with a lower level of crosslinking, especially for low $T$ and large $\lambda$ where nonlinear relaxation processes typical of viscoelastic fluids are active~\cite{wagner1978constitutive}. However, it provides a good description of the nonlinear rheology in the domain of stretch experienced by fibrils before debonding in peeling.

\begin{figure}
    \centering
    \includegraphics[width = 9cm]{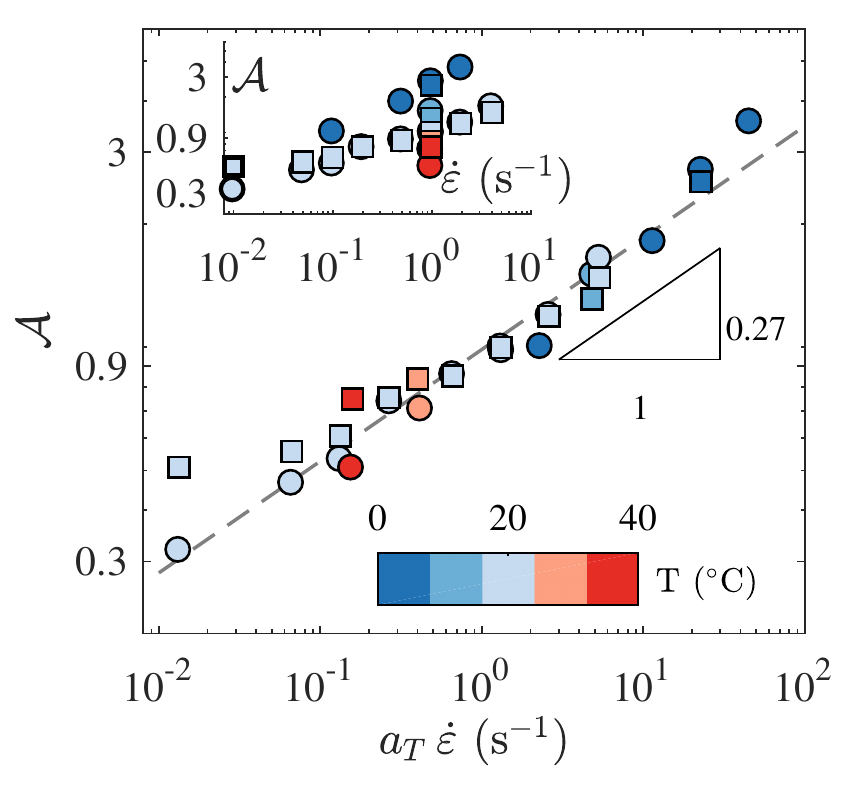}
    \caption{The rescaling factor $\mathcal{A}$ for both polymers is obtained from a fitting procedure over the entire range of strain rates for various temperatures in the range $0$ to $40$\degC (colorbar). $\mathcal{A}(\deps,T)$ depends both on $\deps$ and $T$ (inset:raw data), but can be collapsed to  $\mathcal{A} \sim (a_T\dot{\varepsilon})^{0.27}$ when using the same shift factor $a_T$ measured from the linear viscoelastic behavior.}
    \label{Fig-AShift}
\end{figure}

In the inset of Fig.~\ref{Fig-AShift}, the prefactor $\mathcal{A}$ for polymers A and B is plotted in a log-log scale as a function of $\deps$ for the range of temperatures for which we have performed nonlinear tests. In the main graphic of Fig.~\ref{Fig-AShift}, we show that the data can be collapsed onto a master curve when plotted against $a_T \deps$ where $a_T$ is the shift factor determined from linear rheology. Remarkably, the linear and nonlinear rheology are thus  found to share the same TTS. Moreover, for both polymers, we find that $\mathcal{A} \sim (a_T \deps)^{0.27}$ for $\deps > 10^{-2}$\,s$^{-1}$. Deviations are observed at smaller $\deps$ but, as it will be shown later, this occurs outside the range of strain rates experienced by the fibrils during our peeling tests. By construction, the same prefactor $\mathcal{A}$ can be used to rescale the true stress curves $\sigma(\varepsilon)$, yielding :
\begin{equation}
\sigma(\varepsilon,\deps,T) = \mathcal{A}(a_T\deps) \cdot \sigma^{ref}(\varepsilon)
\label{eq:Afactor}
\end{equation}
which is compatible with  a Rivlin-Sawyers type of nonlinear constitutive law~\cite{bird1987dynamics}. Since Eq.~\ref{eq:Afactor} is also valid in the small strain regime where we have $\sigma \sim G(t) \cdot \eps (t)$, we obtain that $\mathcal{A}(a_T \deps) \sim G(a_T \deps) \sim (a_T \deps)^{\beta}$ where we have used the Cox-Merz rule ($\deps \sim \omega \sim 1/t$). The exponent $\beta = 0.23$ measured by linear rheology (cf.~Fig.~\ref{Fig-LinRheol}(e)) is found to be in reasonable agreement with the 0.27 exponent obtained for nonlinear rheology. This strongly indicates that most of the rate and temperature dependence of both polymers is encapsulated in a single linear rheology function $\mathcal{A}$, in the range relevant for the measured peeling curves.

\begin{figure}
    \centering
    \includegraphics[width = 14cm]{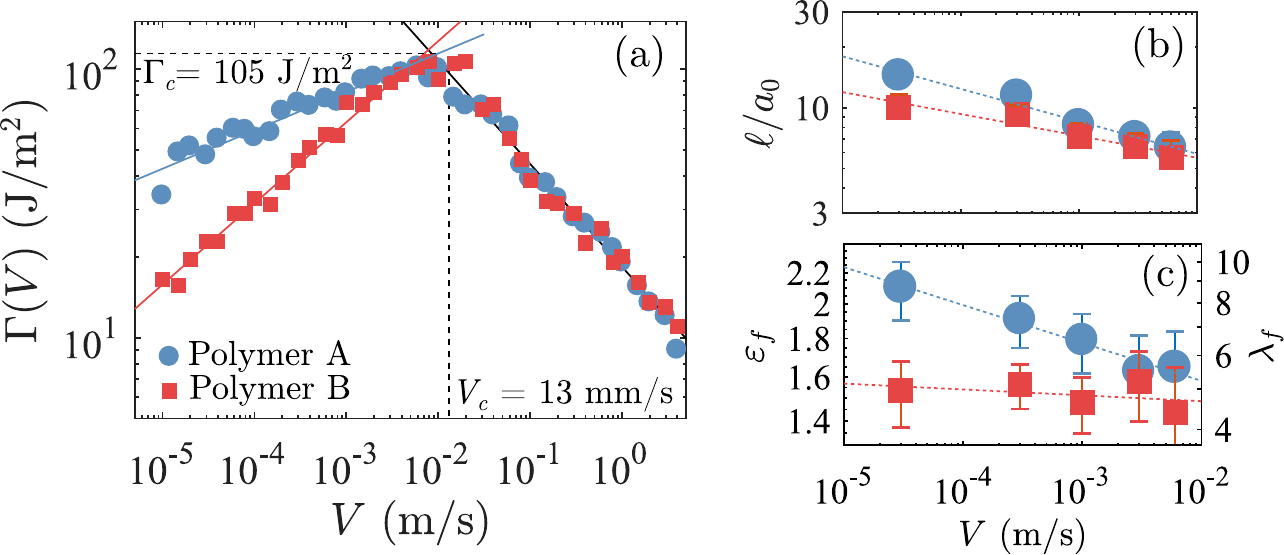}
    \caption{(a) Measured adherence energy $\Gamma(V)$ for polymer A and B. Above a characteristic peeling rate $V_c$, a stick-slip instability develops for both polymers. Data from Villey~\textit{et al.}~\cite{Villey2015}. (b) The reduced lateral size  decreases with speed as $\ell/a_0\sim V^{-p}$ with $p =0.16$ (resp. $p = 0.10$) for polymer A (resp. B). (c) The final strain $\varepsilon_f = \ln(\lambda_f)$ is described by $\varepsilon_f \sim V^{-q}$ where $q =0.05$ (resp. $q = 0.007$) for polymer A (resp. B).
    }
    \label{Fig-CohesiveZone}
\end{figure}

Next, we focus on the characterization of the cohesive zone during steady-state peeling using an instrumented peel test~\cite{Villey2015,villey2017situ} at a 90$^{\circ}$ angle and $T = 23$\degC. From images of the fibrillated debonding region, we measured both the extension $\ell$ of the stretched region and the maximum fibril length $a_f$ at debonding (see Fig.~\ref{Fig-Setups}(d)). In Fig.~\ref{Fig-CohesiveZone}(a), we reproduced the adherence curves as a function of the peeling rate for polymer A and B from data collected in Villey~\textit{et al.}~\cite{Villey2015}. As mentioned in the introduction, polymer B which has a more pronounced strain hardening presents a lower adherence than polymer A but a stronger dependence with the peeling rate. The adherence curves for the two polymers are distinct until reaching a critical velocity $V_c = 13\,$mm/s above which they exhibit an identical trend characterized by a negative slope. The region $V>V_c$ corresponds to an unsteady peeling caused by a stick-slip instability~\cite{Cortet2013}. However, in the present manuscript, we focus on the steady-state peeling regime with a well-defined peeling rate. In this regime, the normalized lateral size $\ell/a_0$ shown in Fig.~\ref{Fig-CohesiveZone}(b) is found to decrease as a power law $\ell \sim V^{-p}$ with an exponent $p \simeq 0.13$ weakly dependent on the crosslinking level. The characteristic timescale for stretching the fibrils $t_f = \ell/V$ is in the range $0.01-20$\,s a regime where linear rheological measurements show no difference between polymers A and B (see  Fig.~\ref{Fig-LinRheol}). The maximal fibril stretch $\lambda_f =a_f/a_0$ plotted in Fig.~\ref{Fig-CohesiveZone}(c) shows that fibrils reach the strain hardening regime before debonding from the substrate (see Fig.~\ref{Fig-NonLinExt}). Interestingly, the maximum stretch increases upon decreasing $V$, indicating that the nonlinear viscoelastic response dominates even in the lower speed limit. The signature of material nonlinearities at low speed is also demonstrated by the clear increase of $\lambda_f$ when the crosslinking level is lowered. A fit of the data with a power law yields $\varepsilon_f = \ln(\lambda_f) \sim V^{-q}$ with $q = 0.05$ (resp.\ $q = 0.007$) for polymer A (resp.\ B). Furthermore, we evaluate the average strain rate $\deps_a$ experienced by a fibril as $\deps_a \approx \eps_f V/\ell \propto V^{1+m}$ where $m = p - q \ll 1$. For the measured peeling rates in the range $10^{-5}- 10^{-2}$\, m/s, the equivalent strain rate is in the range of $5\cdot10^{-2}  -  10^{2}$\,s$^{-1}$ which is overlapped by the rheological measurements.

\begin{figure*}
    \centering
        \includegraphics[width=10cm]{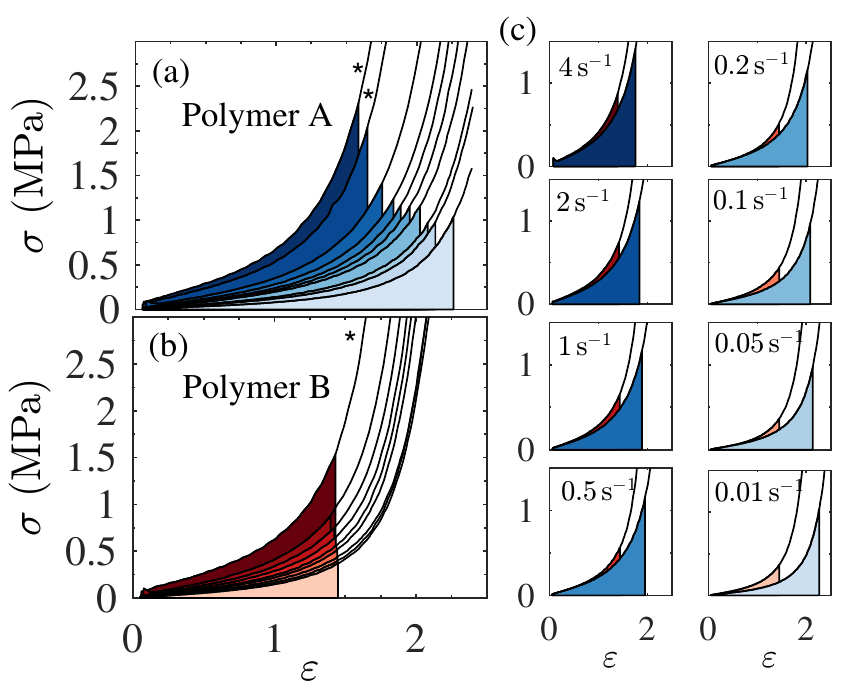}
    \caption{Evolution of the dissipated energy (proportional to the shaded areas) with $a_T\cdot\deps$ (shifted at a common temperature $T = 20$\degC), for (a) polymer A and (b) polymer B. Upper bounds for the integral $\eps_f = \ln \lambda_f$ are obtained from data given in Fig.~\ref{Fig-CohesiveZone}(c). Curves with a star label are measured at $0$\degC\, and rescaled.  (c) Comparison of the stress-strain curves and debonding criteria for polymer A and B  for the same $\deps$ at $T = 20$\degC.}
    \label{Fig-ModelComp}
\end{figure*}

We are now in a position to propose a model to predict the adherence energy by combining the nonlinear material behavior and the measured characteristics of the fibrillated debonding region. Since the cavitation of the adhesive locally relaxes the effect of confinement~\cite{chikina2000cavitation}, we initially make the assumption that the large stretch of the fibrils can be treated as uniaxial. The adherence energy can thus be modeled as:
\begin{equation}
    \Gamma^c(V) = k\cdot a_0 \int_0^{\eps_f(\deps_a)} \sigma(\eps',\deps_a)d\eps'
    \label{Eq_GeneralModel}
\end{equation}
where $\sigma(\varepsilon,\deps_a)$ is provided by the elongational measurements corresponding to the average strain rate $\deps_a$ and $k$ is an adjustable factor compensating the crude approximation of uniaxial extension of fibrils, while
the more complex fibril drawing condition will be investigated  in  future  work. In the present implementation $\varepsilon_f(V)$ and $\ell(V)$ are measured quantities, since a model able to capture their functional dependence with the peeling speed has yet to be developed.

\begin{figure}
    \centering
    \includegraphics[width=10cm]{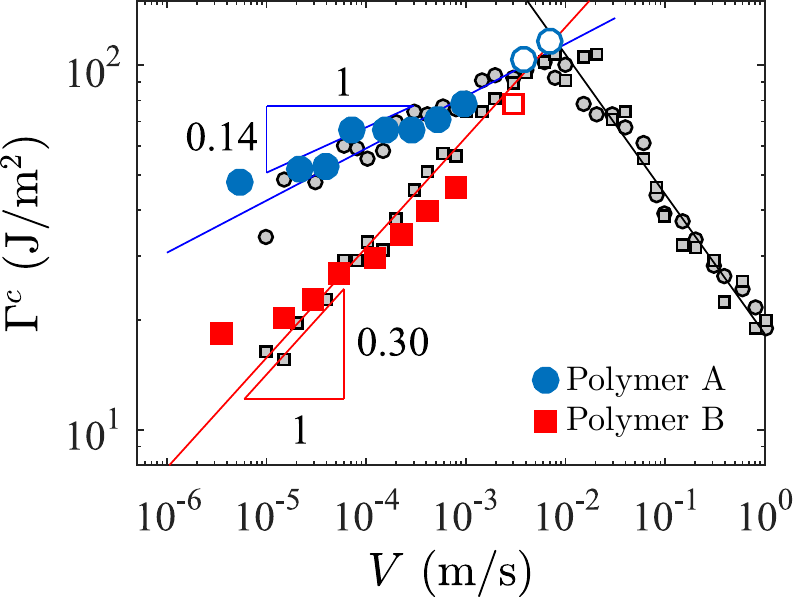}
    \caption{Calculated adherence curves $\Gamma^c$ given by 
    Eq.~\ref{Eq_GeneralModel} superposed with peeling data from Villey~\textit{et al.}~\cite{Villey2015}  (grey symbols), already shown in Fig.~\ref{Fig-CohesiveZone}(a). Using a numerical prefactor $k = 5$ quantitative agreement is obtained. Empty symbols are calculated from measurements taken at $0$\degC.}
    \label{Fig-GammaV}
\end{figure}

In Fig~\ref{Fig-ModelComp}, we replot the $\sigma(\eps,\deps)$ curves for polymer A  and B in terms of true stress and strain varying the strain rate at $T=20$\degC, unless otherwise stated in caption. The area under the curves is proportional to $\Gamma^c/a_0$. In Fig~\ref{Fig-GammaV}, we plot the model prediction $\Gamma^c$ as a function of the peeling rate $V$ according to Eq.~\ref{Eq_GeneralModel}, along with the steady-state peeling data from our previous work \cite{Villey2015}. The two datasets are in excellent agreement if the value of the peel prediction is multiplied by a dimensionless prefactor $k = 5$ demonstrating that Eq.~\ref{Eq_GeneralModel} captures the clear difference in the adherence energy curves $\Gamma(V)$ between polymer A and B both in terms of relative values and slope, while a model based on linear rheology alone cannot explain this difference. The dimensionless prefactor $k$, can be interpreted as the sign that the fiber drawing process from the bulk adhesive layer is affected by a higher stress triaxiality than uniaxial extension tests, which will be the focus of further investigation.

The direct experimental access to the model parameters provides interesting insights on the physical mechanisms at the origin of the adherence curves in peeling. When focusing on the fibril debonding data in Fig.~\ref{Fig-ModelComp}(a), we clearly see that the debonding criterion cannot be simply expressed in terms of a critical stress or strain \cite{Hata1972,Yarusso1999}. However, for the more crosslinked polymer B, the debonding criterion is closer to a critical strain condition. In Fig.~\ref{Fig-ModelComp}(c), we compare the stress curves for polymers A and B at specific values of $\deps$. We observe that although the stress level for B is always higher than for A, the overall dissipation, hence $\Gamma^c$, is larger for A due to  the larger values of $\eps_f$. Moreover, the observation that the two polymers present a different rate dependence of $\eps_f$ with $\deps$ explains why at higher $\deps$ the peeling energy $\Gamma^c$ for both polymers tends to get closer. Therefore, we demonstrate that both the increase in the stress level  with $\deps$, and the decrease of $\eps_f$ \ are crucial to determine  the general trend of $\Gamma^c(V)$, making it difficult to derive {\it a priori} predictions unless we identify a sound fibril debonding criterion.

We can now discuss the link between our model and previous linear models. Taking advantages of the separability between strain rate dependence and strain dependence observed for our polymers (see Eq.~\ref{eq:Afactor}), Eq.~\ref{Eq_GeneralModel} can be further simplified as :
\begin{equation}
    \Gamma^c(V) \propto \mathcal{A}(a_T \deps_a) \cdot \int_0^{\eps_f(\deps_a)} \sigma^{ref}(\eps)d\eps
    \label{Eq:separableGamma}
\end{equation}
which more clearly reveals two contributions to $\Gamma^c(V)$ : 1- a rate-dependent linear viscoelastic factor, and 2- a nonlinear factor whose rate-dependence originates from the debonding criterion. If we first consider polymer B alone, for which $\eps_f$ is essentially rate independent, the dependence on rate of the adherence curve $\Gamma^c(V) \propto V^{n}$ is completely determined by linear rheology and the power law exponent of $n = 0.30$ is quite close to the measured values of $\beta \sim 0.23$ to $0.27$. This result can explain why several authors have previously observed the TTS superposition for the adherence curves and the correlations between the $n$ and $\beta$ exponents. However, when considering the behaviour of polymer A, which has a less elastic character, the maximum fibril stretch at debonding $\eps_f$ becomes significantly dependent on the peel rate and the competition between the increase in stress and the decrease in maximum stretch affects the $\Gamma^c(V)$ curve in a deeply nonlinear manner. It is worth noting that even in the low speed regime, where LEFM based approaches are generally considered to be more appropriate, extremely soft materials, such as polymer A, are shown to be stretched increasingly outside their linear viscoelastic regime, thus precluding the use of linear models. 

\section{Conclusion}
We have shown that the peel force of model soft viscoelastic PSAs can be quantitatively modeled for a range of strain rates with a single adjustable dimensionless prefactor based on experimentally measured values of the maximum stretch in the fibrils at the detachment point. For more crosslinked adhesives the stress can be separated into the product of a strain dependent term that characterizes the strain softening and hardening behaviour of the adhesive, and a strain rate dependent term, which characterizes the molecular friction and is dependent on the glass transition temperature of the adhesive. However, for weakly crosslinked adhesives the two factors are no longer easily separable and the functional form of the strain hardening depends on the strain rate especially for very large strains.

A detailed modeling of both the relevant stress triaxiality during the fibril drawing process and of the fibril debonding criterion are certainly the next step, but we believe that the present model provides a clear understanding of the ingredients needed to understand the adherence energy in the peeling of very soft and dissipative materials.

\end{document}